\documentstyle[12pt]{article}

\def\theorem #1. #2\par{\medbreak
  \noindent{\tt {\bf Theorem #1.}\enspace}{\sl#2\par}%
  \ifdim\lastskip<\medskipamount \removelastskip\penalty55\medskip\fi}

\def\lemma #1. #2\par{\medbreak
  \noindent{\tt {\bf Lemma #1.}\enspace}{\sl#2\par}%
  \ifdim\lastskip<\medskipamount \removelastskip\penalty55\medskip\fi}

\def\proof{\medbreak\noindent{\bf Proof}}

\def\det{{\rm det}}
\def\diag{{\rm diag}}
\def\D{{\cal D}}
\def\M{{\cal M}}
\def\I{{\cal I}}
\def\C{{\cal C}}
\def\L{{\cal L}}
\def\V{{\cal V}}
\def\W{{\cal W}}
\def\Tr{{\rm Tr}}
\def\Spin{{\rm Spin}}

\begin{document}

\title{Dirac-type tensor equations on a parallelisable manyfolds}

\author{N.G.Marchuk \thanks{Research supported by the Russian Foundation
for Basic Research grants 00-01-00224,  00-15-96073.}}

\maketitle

\begin{abstract}
The goal of this work is to extend Dirac-type tensor equations
to a curved space. We take four 1-forms (a tetrad) as a unique
structure, which determines a geometry of space-time.
\end{abstract}


\vskip 1cm

Steklov Mathematical Institute, Gubkina st.8, Moscow 119991, Russia

nmarchuk@mi.ras.ru,   www.orc.ru/\~{}nmarchuk

\bigskip

In \cite{nona1} we suggest to take the differential forms
$H,I,K$ (see item 14 of sect. 1 in \cite{nona}) as an additional
structure on pseudo-Riemannian space. In the current paper,
developing this idea, we take four 1-forms $e^a$ (a tetrad) as
a unique structure, which determines a geometry of
space-time. A metric tensor is expressed via the tetrad.
Hence we arrive at  a geometry, which was
considered by many authors (see, for example,
M\o ller \cite{M}) as a mathematical model of physical
space-time and gravity (according to the Theory of General
Relativity the gravity is identified with the metric
tensor).

The goal of our work, begining at
\cite{nona},\cite{nona1}, is
to extend Dirac-type tensor equations (see \cite{dtt}) to a
curved space.


\section{A pseudo-Riemannian space}
Let $\M$ be a four dimensional differentiable manyfolds with
local coordinates $x^\mu,\, \mu=0,1,2,3$ and with a metric
tensor $g_{\mu\nu}=g_{\nu\mu}$ such that $g_{00}>0$,
$g=\det\|g_{\mu\nu}\|<0$ and the signature of matrix
$\|g_{\mu\nu}\|$ is equal to $-2$. The full set of
$\{\M,g_{\mu\nu}\}$  is called {\it a pseudo-Riemannian
space} and is denoted by $\V$. The metric tensor defines the Levi-Civita
connection, the curvature tensor, the Ricci tensor, and the
scalar curvature
\begin{eqnarray}
{\Gamma_{\mu\nu}}^\lambda&=&
\frac{1}{2}g^{\lambda\kappa}(\partial_\mu g_{\nu\kappa}+
\partial_{\nu}g_{\mu\kappa}-\partial_\kappa g_{\mu\nu}),
\label{Levi-Civita}\\
{R}_{\lambda\mu\nu}{}^\kappa&=&
\partial_\mu {\Gamma}_{\nu\lambda}{}^\kappa-\partial_\nu
{\Gamma}_{\mu\lambda}{}^\kappa+
{\Gamma}_{\mu\eta}{}^\kappa{\Gamma}_{\nu\lambda}{}^\eta-
{\Gamma}_{\nu\eta}{}^\kappa{\Gamma}_{\mu\lambda}{}^\eta,
\label{curv}\\
R_{\nu\rho}&=&{R^\mu}_{\nu\mu\rho},
\label{Ricci}\\
R&=&g^{\rho\nu}R_{\rho\nu}
\label{scalar-curv}
\end{eqnarray}
with symmetries
$$
{\Gamma_{\mu\nu}}^\lambda={\Gamma_{\nu\mu}}^\lambda,\quad
R_{\mu\nu\lambda\rho}=R_{\lambda\rho\mu\nu}=R_{[\mu\nu]\lambda\rho},\quad
R_{\mu[\nu\lambda\rho]}=0,\quad
R_{\nu\rho}=R_{\rho\nu}
$$
Let $\top^p_q$ be the set of all tensor fields of rank $(q,p)$ on $\V$. The
covariant derivatives $\nabla_\mu\,:\,\top^p_q\to\top^p_{q+1}$ are defined
via the Levi-Civita connection by the following rules:
\medskip

\noindent 1. If $t=t(x),\ x\in\V$ is a scalar function, then
$$
\nabla_\mu t=\partial_\mu t.
$$

\noindent 2. If $t^\nu\in\top^1$, then
$$
\nabla_\mu t^\nu\equiv t^\nu_{;\mu}=\partial_\mu t^\nu +
{\Gamma_{\lambda\mu}}^\nu t^\lambda.
$$

\noindent 3. If $t_\nu\in\top_1$, then
$$
\nabla_\mu t_\nu\equiv t_{\nu;\mu}=\partial_\mu t_\nu -
{\Gamma_{\nu\mu}}^\lambda t_\lambda.
$$

\noindent 4. If $u=u^{\nu_1\ldots \nu_k}_{\lambda_1\ldots \lambda_l}\in\top^k_l$,
$v=v^{\mu_1\ldots \mu_r}_{\rho_1\ldots \rho_s}\in\top^r_s$, then
$$
\nabla_\mu(u\otimes v)=(\nabla_\mu u)\otimes v + u\otimes\nabla_\mu v.
$$
\medskip

With the aid of these rules it is easy to calculate covariant derivatives of
arbitrary rank tensor fields. Also we can check the formulas
\begin{eqnarray*}
&&\nabla_\mu g_{\alpha\beta}=0,\quad
\nabla_\mu g^{\alpha\beta}=0,\quad
\nabla_\mu \delta_\alpha^\beta=0,\\
&&\nabla_\alpha(R^{\alpha\beta}-\frac{1}{2}R g^{\alpha\beta})=0,\\
&&(\nabla_\mu \nabla_\nu-\nabla_\nu \nabla_\mu)a_\rho=
{R^\lambda}_{\rho\mu\nu}a_\lambda,
\end{eqnarray*}
for any $a_\rho\in\top_1$.


\section{A parallelisable manyfolds}
An $n$-dimensional differentiable manifolds is called {\it parallelisable}
if there exist $n$ linear independent vector or covector fields on it.
Let $\M$ be a four dimensional parallelisable manifolds with local
coordinates $x=(x^\mu)$ and
$$
{e_\mu}^a={e_\mu}^a(x),\quad\,a=0,1,2,3
$$
be four covector fields on $\M$. This set of four covectors are called
{\it a tetrad}. The full set $\{\M,{e_\mu}^a\}$
is denoted by $\W$.
Here and in what follows we use greek indices as tensorial indices
and latin indices as nontensorial (tetrad) indices, which enumerate
covectors.

Let us take the Minkowski matrix
$$
\eta_{ab}=\eta^{ab}=\diag(1,-1,-1,-1).
$$
Then we can define a metric tensor
\begin{equation}
g_{\mu\nu}={e_\mu}^a {e_\nu}^b \eta_{ab},
\label{metric}
\end{equation}
such that
$$
g_{\mu\nu}=g_{\nu\mu},\quad g_{00}>0,\quad g=\det\|g_{\mu\nu}\|<0
$$
and the signature of the matrix $\|g_{\mu\nu}\|$ is equal to $-2$.
The Levi-Civita connection ${\Gamma_{\mu\nu}}^\lambda$, the curvature
tensor ${R^\mu}_{\nu\lambda\rho}$, the Ricci tensor $R_{\nu\rho}$, the
scalar curvature $R$, and the covariant derivatives $\nabla_\mu$ are
defined via $g_{\mu\nu}$ as in sect. 1.
All constructions of \cite{nona} (the Clifford product of differential
forms, the Spin(1,3) group, Upsilon derivatives $\Upsilon_\mu$, etc.) are
valid in the parallelisable manyfolds $\W$.

We raise and lower latin indices with the aid of the matrix $\eta_{ab}=\eta^{ab}$
and greek indices with the aid of the metric tensor $g_{\mu\nu}$
$$
e^{\nu a}=g^{\mu\nu}{e_\mu}^a,\quad e_{\mu a}=\eta_{ab} {e_\mu}^b.
$$
If we take 1-forms $e^a,e_a\in\Lambda_1$
$$
e^a={e_\mu}^a dx^\mu,\quad
e_a=\eta_{ab}e^b,
$$
then we see that
\begin{equation}
e^a e^b+e^b e^a=2\eta^{ab}.
\label{cliff}
\end{equation}
Indeed,
\begin{eqnarray*}
e^a e^b+e^b e^a &=&e_\mu{}^a dx^\mu e_\nu{}^b dx^\nu+e_\nu{}^b dx^\nu e_\mu{}^a dx^\mu=\\
&=& e_\mu{}^a e_\nu{}^b(dx^\mu dx^\nu+dx^\nu dx^\mu)= e_\mu{}^a e_\nu{}^b 2 g^{\mu\nu}=
2\eta^{ab}.
\end{eqnarray*}

The transformation
\begin{equation}
e^a\,\to\,\check{e}^a=S^{-1}e^a S,
\label{tetrad:rotation}
\end{equation}
where $S\in\Spin(1,3)$, is called {\it a Lorentz rotation of the tetrad}.
Evidently, formula (\ref{cliff})
is invariant under Lorentz rotations of the tetrad, i.e.,
$$
e^a e^b+e^b e^a=2\eta^{ab} \quad \leftrightarrow\quad
\check{e}^a \check{e}^b+\check{e}^b \check{e}^a=2\eta^{ab}.
$$
In the sequel we use the following lemma.

\lemma.
$$
e^a U e_a=\cases{ 4U & for $U\in\Lambda_0\top^p_q$,\cr
-2U & for $U\in\Lambda_1\top^p_q$,\cr
0 & for $U\in\Lambda_2\top^p_q$,\cr
2U & for $U\in\Lambda_3\top^p_q$,\cr
-4U & for $U\in\Lambda_4\top^p_q$.
}
$$
\par

The proof in by direct calculation.
\medskip

Let us take a tensor $B_\mu\in\Lambda_2\top_1$
\begin{equation}
B_\mu=-\frac{1}{4}e^a\wedge\Upsilon_\mu e_a.
\label{Bmu}
\end{equation}

\theorem. Under the Lorentz rotation of tetrad (\ref{tetrad:rotation})
the tensor $B_\mu$ transforms as
$$
B_\mu\to \check{B}_\mu=S^{-1}B_\mu S-S^{-1}\Upsilon_\mu S.
$$
\par

\proof. We have
$
B_\mu=-\frac{1}{4}e^a\wedge\Upsilon_\mu e_a=-\frac{1}{4}e^a\Upsilon_\mu e_a.
$
Therefore
\medskip

$-4\check{B}_\mu=\check{e}^a\Upsilon_\mu \check{e}_a=
S^{-1}e^a S\Upsilon_\mu(S^{-1}e_a S)=
S^{-1}e^a S(\Upsilon_\mu S^{-1})e_a S+S^{-1}e^a \Upsilon_\mu e_a S+
S^{-1}e^a e_a\Upsilon_\mu S=
-4S^{-1}B_\mu S+4S^{-1}\Upsilon_\mu S+S^{-1}e^a S(\Upsilon_\mu S^{-1})e_a S$.
\medskip

Here we use the formula $e^a e_a=4$ from the Lemma. It can be checked that
$S\Upsilon_\mu S^{-1}\in\Lambda_2\top_1$. Consequently from the Lemma we get that
$$
e^a S(\Upsilon_\mu S^{-1})e_a=0.
$$
These completes the proof.\par

Note that the set of 2-forms $\Lambda_2$ can be considered as the real
Lie algebra of the Lie group $\Spin(1,3)$. Hence $B_\mu$ belong to this Lie algebra.
$B_\mu$ is a tensor with respect to changes of coordinates. But, according to the Theorem,
under Lorentz rotations of the tetrad $B_\mu$ transforms as a connection.

Now we may define operators
$\D_\mu=\Upsilon_\mu-[B_\mu,\,\cdot\,]$ acting on tensors
from $\Lambda\top^p_q$ and such that
$$
\D_\mu e^a=0,\quad\D_\mu e_a=0,\quad
\D_\mu(UV)=(\D_\mu U)V+U\D_\mu V,\quad
\D_\mu \D_\nu-\D_\nu\D_\mu=0.
$$
Consider the tensor from $\Lambda_2\top_2$
$$
\frac{1}{2}C_{\mu\nu}=\D_\mu B_\nu-D_\nu B_\mu+[B_\mu,B_\nu].
$$
It can be shown that
$$
C_{\mu\nu}=\frac{1}{2}R_{\mu\nu\alpha\beta}\,dx^\alpha\wedge dx^\beta.
$$
In \cite{nona} (see item 14) of sect.1) we define differential forms
$H\in\Lambda_1$; $I,K\in\Lambda_2$; $\ell\in\Lambda_4$, which we call
{\it secondary generators of $\Lambda$}. These differential forms are
connected with the tetrad $e^a$ by the following formulas:
\begin{eqnarray*}
&&H=e^0,\quad I=-e^1 e^2,\quad K=-e^1 e^3,\quad \ell=e^0 e^1 e^2 e^3,\\
&&e^0=H,\quad e^1=IK\ell H,\quad e^2=K\ell H,\quad e^3=-I\ell H.
\end{eqnarray*}
The formula for $B_\mu$ from \cite{nona1}
\begin{eqnarray*}
B_\mu&=&-\frac{3}{8}H\Upsilon_\mu H+\frac{1}{4}(I\Upsilon_\mu I+K\Upsilon_\mu K)\\
&&+\frac{1}{8}H(I\Upsilon_\mu I+K\Upsilon_\mu K)H-\frac{1}{8}IKH\Upsilon_\mu H\,KI-
\frac{1}{8}(KI\Upsilon_\mu I\,K+IK\Upsilon_\mu K\,I)
\end{eqnarray*}
is equivalent to formula (\ref{Bmu}).


\section{Lagrangians and main equations}
Consider the invariant
$$
\L_2=R+4\Tr(\delta B),
$$
where $R$ is the scalar curvature,
$B=dx^\mu B_\mu$, and the codifferential $\delta\,:\,\Lambda_k\to\Lambda_{k-1}$
was defined in \cite{Cim1}.
It can be checked that the invariant $\L_2$ doesn't depend on second
derivatives of tetrad components ${e_\mu}^a$. Variating the Lagrangian
$\L_2$ with respect to the components of metric tensor $g_{\mu\nu}$,
we get the Einstein tensor
$$
\epsilon\sqrt{-g}(R^{\mu\nu}-\frac{1}{2}g^{\mu\nu}R)=
\frac{\partial(\sqrt{-g}\L_2)}{\partial g_{\mu\nu}}-
\partial_\rho\frac{\partial(\sqrt{-g}\L_2)}{\partial g_{\mu\nu,\rho}},
$$
where $g_{\mu\nu,\rho}=\partial_\rho g_{\mu\nu}$, $\epsilon=1$ for $\mu=\nu$
and $\epsilon=2$ for $\mu\neq\nu$. Note that we can easily calculate the
partial derivatives $\frac{\partial{e_\mu}^a}{\partial g_{\alpha\beta}}$,
$\frac{\partial{e_{\mu,\rho}}^a}{\partial g_{\alpha\beta,\rho}}$ using
formulas
\begin{eqnarray*}
g_{\alpha\beta}&=&{e_\alpha}^a {e_\beta}^b \eta_{ab},\\
\frac{\partial g_{\alpha\beta}}{\partial{e_\mu}^a}&=&
\frac{\partial g_{\alpha\beta,\rho}}{\partial{e_{\mu,\rho}}^a}=
\delta^\mu_\alpha e_{\beta a}+\delta^\mu_\beta e_{\alpha a},\\
\frac{\partial{e_\mu}^a}{\partial g_{\alpha\beta}}&=&
\frac{\partial{e_{\mu,\rho}}^a}{\partial g_{\alpha\beta,\rho}}=
\frac{1}{\delta^\mu_\alpha e_{\beta a}+\delta^\mu_\beta e_{\alpha a}},
\end{eqnarray*}
where $\rho=0,1,2,3$.
Finally, we can take the Lagrangian
\begin{eqnarray}
\L&=&\L_0+\L_1+\L_2\label{L}\\
&=&2\Tr(e^0(\Psi^*P+P^*\Psi))+\frac{1}{4}\Tr(F_{\mu\nu}F^{\mu\nu})+
(R+4\Tr(\delta B)),\nonumber
\end{eqnarray}
where
\begin{eqnarray*}
P&=&dx^\mu(\D_\mu\Psi+\Psi A_\mu+B_\mu\Psi)N-m\Psi E,\\
F_{\mu\nu}&=&\D_\mu A_\nu-\D_\nu A_\mu-[A_\mu,A_\nu].
\end{eqnarray*}
and the Lagrangians $\L_0,\L_1$ were defined in sect.6 of \cite{nona}.
Variating the Lagrangian $\L$ with respect to components of $\Psi,A_\mu$
and w.r.t. components of metric tensor, we arrive at the system of equations
\begin{eqnarray}
&&dx^\mu(\D_\mu\Psi+\Psi A_\mu+B_\mu\Psi)N-m\Psi E=0,\nonumber\\
&&\frac{1}{\sqrt{-g}}\D_\mu(\sqrt{-g}F^{\mu\nu})-[A_\mu,F^{\mu\nu}]=J^\nu,
\label{main}\\
&&R^{\mu\nu}-\frac{1}{2}Rg^{\mu\nu}=-T^{\mu\nu},\nonumber
\end{eqnarray}
where $J^\nu$ are defined in (27) of \cite{nona} and $T^{\mu\nu}$ is the
energy-momentum tensor
$$
\epsilon\sqrt{-g}\,T^{\mu\nu}=
\frac{\partial(\sqrt{-g}(\L_0+\L_1))}{\partial g_{\mu\nu}}-
\partial_\rho\frac{\partial(\sqrt{-g}(\L_0+\L_1))}{\partial g_{\mu\nu,\rho}}.
$$
Let us remark that we may insert two constants into Lagrangian
$\L=\L_0+c_1 \L_1+c_2 \L_2$ in (\ref{L}) and into equations (\ref{main})
respectively. Constants $c_1,c_2$ depend on physical units and on
experimental data.


\section{Comparing the Dirac equation with the Dirac-type tensor
equation}
It is well known that the Dirac equation for the electron has the following
form in a curved space (see, for example, \cite{Weyl}):
\begin{equation}
\gamma^c e^\mu{}_c(\partial_\mu + ia_\mu-
\omega_{\mu ab}\frac{1}{4}[\gamma^a,\gamma^b])\psi+im\psi=0,
\label{standard:Dirac}
\end{equation}
where $\gamma^a$ are complex valued $4\!\times\!4$-matrices with the property
$\gamma^a\gamma^b+\gamma^b\gamma^a=2\eta^{ab}{\bf 1}$, ${\bf 1}$ is identity matrix,
and $\omega_{\mu ab}=\omega_{\mu[ab]}$ is a Lorentz connection. Now we show that
the Dirac type tensor equation
\begin{equation}
dx^\mu(\D_\mu\Psi+a_\mu\Psi I+B_\mu\Psi)+m\Psi HI=0
\label{tensor:eq}
\end{equation}
can be written in the same form (\ref{standard:Dirac}).
A method of reduction of (\ref{tensor:eq}) to (\ref{standard:Dirac}) was developed
in \cite{vs} for the case of Minkowski space.

Let us take the idempotent differential form $t=t^2\in\Lambda^\C$
$$
t=\frac{1}{4}(1+H)(1-iI)
$$
and the left ideal
$$
\I(t)=\{Ut\,:\,U\in\Lambda^\C\}\subset\Lambda^\C.
$$
The exterior forms $t_1,\ldots,t_4\in\I(t)$
$$
t_1=t,\quad t_2=Kt,\quad t_3=-I\ell t,\quad t_4=-KI\ell t
$$
are linear independent and they can be considered as basis elements of $\I(t)$.
These differential forms $t_k$ define a map
$\gamma\,:\,\Lambda\top^q_p\to M(4,\C)\top^q_p$ by the formula
$$
U^{\nu_1\ldots\nu_q}_{\mu_1\ldots\mu_p}t_k=
\gamma(U^{\nu_1\ldots\nu_q}_{\mu_1\ldots\mu_p})^n_k t_n,
$$
where $M(4,\C)\top^q_p$ is the set of all rank $(p,q)$ tensors with values
in $4\!\times\!4$ complex matrices and $\gamma(U)^n_k$ is elements of the matrix
$\gamma(U)$ (an upper index enumerates rows and a lower index enumerates columns).
It is easily shown that
$$
\gamma(UV)=\gamma(U)\gamma(V)
$$
for $U\in\Lambda\top^q_q$, $V\in\Lambda\top^s_r$. If we take
$dx^\mu=\delta^\mu_\nu dx^\nu\in\Lambda_1\top^1$, then we get
$$
dx^\mu t_k=\gamma(dx^\mu)^n_k t_n.
$$
Denoting $\gamma^\mu=\gamma(dx^\mu)$, we see that the equality
$dx^\mu dx^\nu+dx^\nu dx^\mu=2g^{\mu\nu}$ leads to the equality
$\gamma^\mu\gamma^\nu+\gamma^\nu\gamma^\mu=2g^{\mu\nu}{\bf 1}$.
Also we have
$$
B_\mu t_p=\gamma(B_\mu)^k_p t_k.
$$
Let us multiply (\ref{tensor:eq}) by $t$. Then
\begin{eqnarray*}
0&=&(dx^\mu(\D_\mu\Psi+a_\mu\Psi I+B_\mu\Psi)+m\Psi HI)t\\
&=&dx^\mu(\D_\mu(\Psi t)+ia_\mu(\Psi t)+B_\mu(\Psi t))+im(\Psi t)\\
&=&dx^\mu(\D_\mu(\psi^k t_k)+ia_\mu(\psi^k t_k)+B_\mu(\psi^p t_p))+im(\psi^n t_n)\\
&=&(dx^\mu t_k)(\partial_\mu\psi^k+ia_\mu\psi^k+\gamma(B_\mu)^k_p\psi^p)+im(\psi^n t_n)\\
&=&((\gamma^\mu)^n_k(\partial_\mu\psi^k+ia_\mu\psi^k+\gamma(B_\mu)^k_p\psi^p)+im\psi^n)t_n.
\end{eqnarray*}
As $t_1,\ldots,t_4$ are linear independent, we see that
$$
(\gamma^\mu)^n_k(\partial_\mu\psi^k+ia_\mu\psi^k+\gamma(B_\mu)^k_p\psi^p)+im\psi^n=0,
\quad n=1,\ldots,4.
$$
These four equations can be written as one equation
\begin{equation}
\gamma^\mu(\partial_\mu+ia_\mu+\gamma(B_\mu))\psi+im\psi=0,
\label{3q}
\end{equation}
where $\psi=(\psi^1\,\ldots\,\psi^4)^T$. We may write $B_\mu\in\Lambda_2\top_1$ as
$$
B_\mu=\frac{1}{2}b_{\mu ab}e^a\wedge e^b=\frac{1}{4}b_{\mu ab}(e^a e^b-e^b e^a),
$$
where $b_{\mu ab}=b_{\mu[ab]}$. This imply that
$$
\gamma(B_\mu)=\frac{1}{4}b_{\mu ab}[\gamma^a,\gamma^b],\quad
\gamma^a=\gamma(e^a),\quad
\gamma^a\gamma^b+\gamma^b\gamma^a=2\eta^{ab}{\bf 1}.
$$
Note that $\gamma^\mu=\gamma^c e^\mu{}_c$ and the eq. (\ref{3q}) can be
written in the form
\begin{equation}
\gamma^c e^\mu{}_c(\partial_\mu+ia_\mu+b_{\mu ab}\frac{1}{4}[\gamma^a,\gamma^b])\psi+im\psi=0.
\label{4q}
\end{equation}
Consequently eqs. (\ref{standard:Dirac}) and (\ref{4q}) are coinside iff a Lorentz
connection is defined by the formula $\omega_{\mu ab}=-b_{\mu ab}$.


\end{document}